\documentclass{article}
\usepackage{amsfonts}
\usepackage{amsmath}

\providecommand{\U}[1]{\protect \rule{.1in}{.1in}}
\RequirePackage{CJK}
\AtBeginDocument{\begin{CJK*}{GBK}{song}\CJKtilde}
\AtEndDocument{\end{CJK*}}

\begin{document}

\title{Partition of unity with mixed quantum states\thanks{%
Project supported by the National Natural Science Foundation of China (Grant
Nos. 1157429, 11775208 and 11264018), Middle-aged and young teachers'
education in fujian province science and technology projects (No. JK
2014053), and the special foundation of Wuyi college young teachers (XQ
201303). }}
\author{Hongyi Fan$^{1},\ $Jun-hua Chen$^{1,2}$, Dehui Zhan$^{3}$, and Liyun
Hu$^{4}$ \\
\leftline{$^{1}${\small Univ Sci \& Technol China, Dept Mat Sci \& Engn, CAS Key Lab Mat Energy Convers, Hefei 230026, P. R. China}}\\
\leftline{$^{2}${\small Univ Sci \& Technol China, Synerget Innovat Ctr Quantum Informat \& Quantum P, Hefei 230026, P. R. China}}\\
\leftline{$^{3}${\small College of Mechanic and Electronic Engineering, Wuyi University, Wuyishan, 354300, P. R. China}}\\
\leftline{$^{4}${\small Center of Quantum Science and Technology, Jinagxi Normal University, Nanchang, 330022, P. R. China}}}
\maketitle

\begin{abstract}
The completeness of quantum state space, is usually expressed as $%
\sum_{m=0}^{\infty }\left\vert m\right\rangle \langle m|=1,$ where $\left\{
\left\vert m\right\rangle \right\} $ is selected set of quantum states
(basis). Density matrix $\left\vert m\right\rangle \langle m|$ describes a
pure quantum state. In this paper, by virtue of the summation method within
the normally ordered product of operators we propose and show that the
completeness relation can also be represented or partitioned in terms of
some mixed states, such as binomial states and negative binomial states.
Thus the view on the structure of Fock space is widen and the connotation of
Fock space is enriched. See the matter in this sight, experimentalists may
have interests to prepare the binomial- and negative binomial states.

\end{abstract}

\section{Introduction}

In quantum mechanics and quantum optics theory, the fundamental concept of
Fock space is frequently used for describing photons' creation and/or
annihilation, and the photon number state is denoted by%
\begin{equation}
\left\vert m\right\rangle =\frac{a^{\dag m}}{\sqrt{m!}}\left\vert
0\right\rangle ,  \label{1}
\end{equation}%
which is a discrete pure state. Here $a^{\dag }$ is bosonic creation operator satisfying the commutative relation $\left[ a,a^{\dag }\right] =1$.
The $\left\vert m\right\rangle $'s form a complete set of Fock space, i.e.,
\begin{equation}
\sum_{m=0}^{\infty }\left\vert m\right\rangle \langle m|=1,  \label{2}
\end{equation}%
in other words, the unity operator can be expanded by the discrete pure
state $\left\vert m\right\rangle \langle m|$. The unity operator in Fock
space can be also expanded by the continuous pure states%
\begin{equation}
\int \frac{d^{2}\alpha }{\pi }\left\vert \alpha \right\rangle \langle \alpha
|=1,  \label{3}
\end{equation}%
where $\left\vert \alpha \right\rangle $ is the coherent states%
\begin{equation}
\left\vert \alpha \right\rangle =e^{\alpha a^{\dag }-\alpha ^{\ast
}a}\left\vert 0\right\rangle .  \label{4}
\end{equation}%
By virtue of the integration within normally product of operators and the
normally ordering form of the vacuum state%
\begin{equation}
\left\vert 0\right\rangle \left\langle 0\right\vert =:e^{-a^{\dag }a}:
\label{5}
\end{equation}%
we can re-express (\ref{3}) as
\begin{equation}
\int \frac{d^{2}\alpha }{\pi }\left\vert \alpha \right\rangle \langle \alpha
|=\int \frac{d^{2}\alpha }{\pi }:e^{-\left( \alpha ^{\ast -}a^{\dag }\right)
\left( \alpha -a\right) }:=1  \label{6}
\end{equation}

We have many different types of quantum light field, which are usually
described by density operators (pure or mixed states) \cite{1}. For
instance, density operator $\rho _{c}$ of the chaotic field is%
\begin{equation}
\rho _{c}=\sum_{m=0}^{\infty }\gamma \left( 1-\gamma \right) ^{m}\left\vert
m\right\rangle \langle m|,  \label{7}
\end{equation}%
An interesting question, which has been overlooked for long, is: can unity
operator be expanded in terms of some density operators which describe mixed
states? If yes, then what are these mixed states? In this paper, we shall
point out that unity operator in Fock space can be expressed in terms of the
binomial state (BS) and the negative binomial state (NBS) respectively.

We shall demonstrate this with the method of summation within the normally
ordered operator because the creation and annihilation operators are
permutable within the normally ordering symbol (denoted by $:$ $:$) \cite%
{2,2a,2b}. Our new partition of unity in Fock space can deepen people's
understanding about the structure of Fock space and enrich the connotation
of Fock space. Once theoretical physicists are certain that some mixed
states can constitute Fock space, experimentalists are challenged to bring
them about.

\section{Unity in Fock space partitioned by the binomial states}

\bigskip Based on the binomial distribution $%
\begin{pmatrix}
n \\
l%
\end{pmatrix}%
\sigma ^{l}\left( 1-\sigma \right) ^{n-l},$ $0<\sigma <1,$ we can construct
the density operator of binomial states of light field
\begin{equation}
\sum_{l=0}^{n}\binom{n}{l}\sigma ^{l}\left( 1-\sigma \right) ^{n-l}|l\rangle
\langle l|\equiv \rho _{n}\left( \sigma \right) ,  \label{8}
\end{equation}%
where $\left\vert l\right\rangle =\frac{a^{\dag l}}{\sqrt{l!}}\left\vert
0\right\rangle $ is the number state and $\mathtt{tr}\rho _{n}\left( \sigma
\right) =1$ is the result of the binomial theorem. Here we should emphasize
that $\rho _{n}\left( \sigma \right) $ denotes a mixed state, sharply
different from the pure state $\sum\limits_{l=0}^{n}\sqrt{\binom{n}{l}\sigma
^{l}\left( 1-\sigma \right) ^{n-l}}|l\rangle \equiv |\psi \rangle $
introduced in Ref.\cite{3}. Using the generating function of Laguerre
polynomials $L_{n}\left( x\right) $,

\begin{equation}
\left( 1-z\right) ^{-1}e^{\frac{z}{z-1}x}=\sum_{n=0}^{\infty }L_{n}\left(
x\right) z^{n},  \label{9}
\end{equation}%
and noticing that operators $a^{\dag }$ and $a$\ are permutable within the
normally ordering symbol $:$ $:$, we can make a decomposition of unity as
the following
\begin{eqnarray}
1 &=&\colon e^{a^{\dag }a}e^{-a^{\dag }a}\colon  \label{10} \\
&=&\sigma \frac{1}{1-\left( 1-\sigma \right) }\colon e^{\frac{1-\sigma }{%
-\sigma }\left( \frac{\sigma }{\sigma -1}a^{\dag }a\right) }e^{-a^{\dag
}a}\colon  \notag \\
&=&\sigma \sum_{n=0}^{\infty }\left( 1-\sigma \right) ^{n}\colon L_{n}\left(
\frac{\sigma }{\sigma -1}a^{\dag }a\right) e^{-a^{\dag }a}\colon ,  \notag
\end{eqnarray}%
where the power series representation of the Laguerre polynomials $%
L_{n}\left( x\right) $ is
\begin{equation}
L_{n}\left( x\right) =\sum_{l=0}^{n}\binom{n}{l}\frac{\left( -\right) ^{l}}{%
l!}x^{l}.  \label{11}
\end{equation}%
Using the normally ordered form of the vacuum projector in Eq.(\ref{5}), we
can rewrite Eq.(\ref{9}) as%
\begin{eqnarray}
1 &=&\sigma \sum_{n=0}^{\infty }\left( 1-\sigma \right) ^{n}\sum_{l=0}^{n}%
\binom{n}{l}\frac{\left( -\right) ^{l}}{l!}\colon \left( \frac{\sigma }{%
\sigma -1}a^{\dag }a\right) ^{l}e^{-a^{\dag }a}\colon  \label{12} \\
&=&\sigma \sum_{n=0}^{\infty }\left( 1-\sigma \right) ^{n}\sum_{l=0}^{n}%
\binom{n}{l}\frac{\left( -\right) ^{l}}{l!}\left( \frac{\sigma }{\sigma -1}%
\right) ^{l}a^{\dag l}|0\rangle \langle 0|a^{l}  \notag \\
&=&\sigma \sum_{n=0}^{\infty }\sum_{l=0}^{n}\binom{n}{l}\sigma ^{l}\left(
1-\sigma \right) ^{n-l}|l\rangle \langle l|.  \notag
\end{eqnarray}%
Thus using Eqs.(\ref{8}), we can write (\ref{12}) in the compact form
\begin{equation}
1=\sigma \sum_{n=0}^{\infty }\rho _{n}\left( \sigma \right) ,  \label{13}
\end{equation}%
which indicates that the unity can be represented or partitioned in terms of
the binomial states. This guides us to theoretically discuss how this mixed
state can be obtained experimentally.

In the next section we will show that when a pure number state $|l\rangle
\langle l|$ passes through an amplitude dissipation channel, the output
state is just the binomial state.

\section{Preparation of the binomial state}

The master equation for amplitude dissipation channel is$^{\cite{FanChen}}$

\begin{equation}
\frac{d}{dt}\rho \left( t\right) =\kappa \left[ 2a\rho a^{\dagger
}-a^{\dagger }a\rho -\rho a^{\dagger }a\right] ,  \label{14}
\end{equation}%
where $\kappa $ is the damping constant. Its solution is%
\begin{equation*}
\rho \left( t\right) =\sum_{n=0}^{\infty }\frac{T^{n}}{n!}\exp \left[
-\kappa ta^{\dagger }a\right] a^{n}\rho _{0}a^{\dagger n}\exp \left[ -\kappa
ta^{\dagger }a\right] ,\text{ \ }T=1-e^{-2\kappa t}.
\end{equation*}%
Taking the initial density operator $\rho _{0}=\left\vert m\right\rangle
\left\langle m\right\vert ,$ and using $a|m\rangle =\sqrt{m}|m-1\rangle $ we
have

\begin{align}
& \rho \left( t\right) =\sum_{n=0}^{\infty }\frac{T^{n}}{n!}\exp \left[
-\kappa ta^{\dagger }a\right] a^{n}\left\vert m\right\rangle \left\langle
m\right\vert a^{\dagger n}\exp \left[ -\kappa ta^{\dagger }a\right]
\label{15} \\
& =\sum_{n=0}^{\infty }\frac{T^{n}}{n!}\exp \left[ -\kappa ta^{\dagger }a%
\right] \frac{m!}{\left( m-n\right) !}\left\vert m-n\right\rangle
\left\langle m-n\right\vert \exp \left[ -\kappa ta^{\dagger }a\right]  \notag
\\
& =\sum_{n=0}^{\infty }T^{n}\left( e^{-2kt}\right) ^{m-n}\binom{m}{n}%
\left\vert m-n\right\rangle \left\langle m-n\right\vert  \notag \\
& =\sum_{n^{\prime }=0}^{m}\binom{m}{m-n^{\prime }}\left( e^{-2kt}\right)
^{n^{\prime }}\left( 1-e^{-2kt}\right) ^{m-n^{\prime }}\colon \frac{%
a^{\dagger n^{\prime }}a^{n^{\prime }}}{n^{\prime }!}e^{-a^{\dagger }a}\colon
\notag \\
& =\sum_{n^{\prime }=0}^{m}\binom{m}{m-n^{\prime }}\left( e^{-2kt}\right)
^{n^{\prime }}\left( 1-e^{-2kt}\right) ^{m-n^{\prime }}\left\vert n^{\prime
}\right\rangle \left\langle n^{\prime }\right\vert ,  \notag
\end{align}%
this is just the binomial state, as defined in (\ref{8}). Thus the binomial
states are experimentally preparable.

\section{Partition of unity in Fock space by negative binomial states}

Next we consider whether the unity of Fock space can be partitioned in terms
of negative binomial states. For this purpose, we employ the following sum
rearranging formula
\begin{equation}
\sum_{n=0}^{\infty }\sum_{l=0}^{n}A_{n-l}B_{l}=\sum_{s=0}^{\infty
}\sum_{m=0}^{\infty }A_{s}B_{m},  \label{26}
\end{equation}%
and rewrite Eq.(\ref{12}) as%
\begin{eqnarray}
1 &=&\sigma \sum_{n=0}^{\infty }\rho _{n}\left( \sigma \right)  \label{27} \\
&=&\sigma \sum_{n=0}^{\infty }\sum_{l=0}^{n}\binom{n}{n-l}\left( 1-\sigma
\right) ^{n-l}\sigma ^{l}|l\rangle \langle l|  \notag \\
&=&\sum_{m=0}^{\infty }\sum_{s=0}^{\infty }\binom{m+s}{s}\left( 1-\sigma
\right) ^{s}\sigma ^{m+1}|m\rangle \langle m|,  \notag
\end{eqnarray}%
where $\binom{m+s}{s}\left( 1-\sigma \right) ^{s}\sigma ^{m+1}$ is the
negative binomial distribution \cite{5,6,7,7a}.

Letting $1-\sigma =\gamma $ in Eq. (\ref{27}), we have

\begin{equation}
1=\sum_{s=0}^{\infty }\sum_{m=0}^{\infty }\binom{m+s}{m}\gamma ^{s}\left(
1-\gamma \right) ^{m+1}|m\rangle \langle m|,  \label{28}
\end{equation}%
By defining

\begin{equation}
\sum_{m=0}^{\infty }\binom{m+s}{m}\gamma ^{s+1}\left( 1-\gamma \right)
^{m}|m\rangle \langle m|\equiv \rho _{s}\left( \gamma \right) ,  \label{29}
\end{equation}%
which is named negative binomial state (it is also a mixed state), and using
the negative binomial theorem

\begin{equation}
\left( 1+x\right) ^{-\left( s+1\right) }=\sum_{m=0}^{\infty }\binom{m+s}{m}%
\left( -x\right) ^{m},  \label{30}
\end{equation}%
we see

\begin{equation}
tr\rho _{s}\left( \gamma \right) =\gamma ^{s+1}\sum_{m=0}^{\infty }\binom{m+s%
}{m}\left( 1-\gamma \right) ^{m}=1.  \label{31}
\end{equation}%
Then the unity decomposition in Eq. (\ref{27}) becomes

\begin{equation}
1=\frac{1-\gamma }{\gamma }\sum_{s=0}^{\infty }\rho _{s}\left( \gamma
\right) .  \label{32}
\end{equation}%
This indicates that unity in Fock space can be partitioned using the
negative binomial states.

Our next task is to see how NBS can be generated theoretically.

\section{Generation of the negative binomial state}

We use $a\left\vert n\right\rangle =\sqrt{n}\left\vert n-1\right\rangle $ and

\begin{equation}
a^{s}\left\vert n\right\rangle =\sqrt{\frac{n!}{\left( n-s\right) !}}%
\left\vert n-s\right\rangle  \label{33}
\end{equation}%
to convert $\rho _{s}$ in (29) to

\begin{eqnarray}
\rho _{s}\left( \gamma \right) &=&\gamma ^{s+1}\sum_{n^{\prime }=s}^{\infty }%
\binom{n^{\prime }}{n^{\prime }-s}\left( 1-\gamma \right) ^{n^{\prime
}-s}\left\vert n^{\prime }-s\right\rangle \left\langle n^{\prime
}-s\right\vert  \label{34} \\
&=&\frac{\gamma ^{s+1}}{\left( 1-\gamma \right) ^{s}s!}\sum_{n=0}^{\infty
}\left( 1-\gamma \right) ^{n}\frac{n!}{\left( n-s\right) !}\left\vert
n-s\right\rangle \left\langle n-s\right\vert  \notag \\
&=&\frac{\gamma ^{s}}{\left( 1-\gamma \right) ^{s}s!}a^{s}\sum_{n=0}^{\infty
}\gamma \left( 1-\gamma \right) ^{n}\left\vert n\right\rangle \left\langle
n\right\vert a^{\dagger s}  \notag \\
&=&\frac{1}{s!\left( n_{c}\right) ^{s}}a^{s}\rho _{c}a^{\dagger s}  \notag
\end{eqnarray}%
where

\begin{equation}
n_{c}\equiv \frac{1-\gamma }{\gamma }.  \label{35}
\end{equation}%
Recalling that
\begin{equation}
\rho _{c}=\sum_{n=0}^{\infty }\gamma \left( 1-\gamma \right) ^{n}\left\vert
n\right\rangle \left\langle n\right\vert =\gamma \left( 1-\gamma \right)
^{a^{\dagger }a}=\gamma e^{a^{\dagger }a\ln \left( 1-\gamma \right) }=\gamma
:e^{-\gamma a^{\dagger }a}:  \label{36}
\end{equation}%
describes a thermo light field (chaotic light), we can prove that $n_{c}$ is
just the average photon number of the chaotic light field. Indeed, using the
photon number operator $N=a^{\dagger }a,$ we have

\begin{eqnarray}
Tr\left( \rho _{c}N\right) &=&\sum_{n=0}^{\infty }\gamma \left( 1-\gamma
\right) ^{n}\left\langle n\right\vert a^{\dagger }a\left\vert n\right\rangle
=\gamma \sum_{n=0}^{\infty }n\left( 1-\gamma \right) ^{n}  \label{37} \\
&=&\frac{1-\gamma }{\gamma }=n_{c}  \notag
\end{eqnarray}%
thus $\gamma =\frac{1}{n_{c}+1}.$ Employing the formula converting an
operator to its anti-normally ordered form \cite{9}, i.e.,
\begin{equation}
\rho (a,a^{\dagger })=\int \frac{d^{2}\beta }{\pi }\vdots \left\langle
-\beta \right\vert \rho (a,a^{\dagger })\left\vert \beta \right\rangle
e^{|\beta |^{2}+\beta ^{\ast }a-\beta a^{\dagger }+a^{\dagger }a}\vdots
\label{38}
\end{equation}%
where the symbol $\vdots \ \vdots $ denotes anti-normally ordering, and $%
\left\vert \beta \right\rangle =\exp [-|\beta |^{2}/2+\beta a^{\dagger
}]\left\vert 0\right\rangle $ is a coherent state, $\left\langle -\beta
\right\vert \left. \beta \right\rangle =e^{-2|\beta |^{2}}$, then $\rho _{c}$
can be put into the following form
\begin{eqnarray}
\rho _{c} &=&\gamma \text{ }\int \frac{d^{2}\beta }{\pi }\vdots \left\langle
-\beta \right\vert \colon e^{-\gamma a^{\dagger }a}\colon \left\vert \beta
\right\rangle e^{|\beta |^{2}+\beta ^{\ast }a-\beta a^{\dagger }+a^{\dagger
}a}\vdots  \label{39} \\
&=&\frac{\gamma }{1-\gamma }\vdots e^{\frac{\gamma }{\gamma -1}aa^{\dagger
}}\vdots .  \notag
\end{eqnarray}%
thus the density operator of negative binomial state in Eq. (\ref{34})
becomes
\begin{equation}
\rho _{s}\left( \gamma \right) =\frac{1}{s!\left( n_{c}\right) ^{s+1}}\vdots
a^{s}e^{\frac{\gamma }{\gamma -1}aa^{\dagger }}a^{\dagger s}\vdots
\label{40}
\end{equation}%
Noticing that operators $a^{\dag }$ and $a$\ are also permutable within the
anti-normally ordering symbol $\vdots \ \vdots ,$ we can perform the
summation over $s$ and again obtain%
\begin{eqnarray}
\sum_{s=0}^{\infty }\rho _{s}\left( \gamma \right) &=&\frac{1}{n_{c}}%
\sum_{s=0}^{\infty }\frac{1}{s!\left( n_{c}\right) ^{s}}\vdots
a^{s}a^{\dagger s}e^{\frac{-1}{n_{c}}aa^{\dagger }}\vdots  \label{41} \\
&=&\frac{1}{n_{c}}\vdots e^{\frac{1}{n_{c}}aa^{\dagger }}e^{\frac{-1}{n_{c}}%
aa^{\dagger }}\vdots =\frac{1}{n_{c}}.  \notag
\end{eqnarray}%
which coincides with Eq.(\ref{32}). Thus the conclusion of Fock space
partitioned by the negative binomial states (\ref{32}) is confirmed. Moreover 
from $\rho _{s}=\frac{1}{s!\left( n_{c}\right) ^{s}}a^{s}\rho
_{c}a^{\dagger s}$ we may predict that the negative binomial state can be
generated by absorbing $s$ photons from the chaotic field $\rho _{c}$, and
this may happen in a light-atom interaction governed by the interacting
Hamiltonian $H=ga^{s}\sigma _{+}+g^{\ast }a^{\dagger s}\sigma _{-},$ here $%
\sigma _{\pm }$ are atom's hopping operators.

\section{Derivation of the generalized negative binomial theorem involving
Laguerre polynomial}

\bigskip As an application of Eq. (\ref{40}) we can use the form of $%
\rho _{s}\left( \gamma \right) $ to derive generalized negative binomial
theorem involving Laguerre polynomial. For this purpose, we need to reform (%
\ref{40}) as in normal ordering. Introducing so-called two-variable Hermite
polynomial through its generating function
\begin{equation}
\sum_{m,n=0}\frac{t^{m}\tau ^{n}}{m!n!}H_{m,n}(x,y)=\exp \left( tx+\tau
y-t\tau \right) ,  \label{42}
\end{equation}%
then%
\begin{align}
H_{m,n}(x,y)& =\frac{\partial ^{n+m}}{\partial t^{m}\partial \tau ^{n}}\exp
\left( tx+\tau y-t\tau \right) |_{t=\tau =0}  \label{43} \\
& =\frac{\partial ^{m}}{\partial t^{m}}e^{tx}\frac{\partial ^{n}}{\partial
\tau ^{n}}\exp \left( \tau \left( y-t\right) \right) |_{t=\tau =0}  \notag \\
& =\frac{\partial ^{m}}{\partial t^{m}}\left[ e^{tx}\left( y-t\right) ^{n}%
\right] |_{t=0}  \notag \\
& =\sum_{l=0}\binom{m}{l}\frac{\partial ^{l}}{\partial t^{l}}\left(
y-t\right) ^{n}\frac{\partial ^{m-l}}{\partial t^{m-l}}e^{tx}|_{t=0}  \notag
\\
& =\sum_{l=0}^{\min (m,n)}\frac{m!n!(-1)^{l}}{l!(m-l)!(n-l)!}x^{m-l}y^{n-l}
\notag
\end{align}%
Comparing it with the Laguerre polynomial in Eq.(\ref{11}), we identify%
\begin{equation}
L_{n}\left( xy\right) =\frac{\left( -1\right) ^{n}}{n!}H_{n,n}\left(
x,y\right) .  \label{44}
\end{equation}%
Then we employ the completeness relation of the coherent state representation

\begin{equation}
\int \frac{d^{2}z}{\pi }\left\vert z\right\rangle \left\langle z\right\vert
=\int \frac{d^{2}z}{\pi }:e^{-|z|^{2}+za^{\dagger }+z^{\ast }a-a^{\dagger
}a}:=1,  \label{45}
\end{equation}%
where $\left\vert z\right\rangle =\exp \left[ -\frac{|z|^{2}}{2}+za^{\dag }%
\right] \left\vert 0\right\rangle $, and the method of integration within
ordered product of operators (IWOP) \cite{2} to derive

\begin{eqnarray}
\vdots e^{\lambda aa^{\dagger }}\vdots &=&\int \frac{d^{2}z}{\pi }e^{\lambda
|z|^{2}}\left\vert z\right\rangle \left\langle z\right\vert =\int \frac{%
d^{2}z}{\pi }e^{\lambda |z|^{2}}e^{-|z|^{2}+z^{\ast }a+za^{\dagger
}-a^{\dagger }a}:  \label{46} \\
&=&1-\lambda ^{-1}:\exp \left[ \frac{-\lambda a^{\dagger }a}{\lambda -1}%
\right] :  \notag
\end{eqnarray}%
Further, using and the generating function of the Laguerre polynomial in (%
\ref{9}) we have

\begin{equation}
\vdots e^{\lambda aa^{\dagger }}\vdots =\colon \sum_{l=0}\lambda
^{l}L_{l}\left( -a^{\dagger }a\right) \colon .  \label{47}
\end{equation}%
It follows from%
\begin{equation}
\int \frac{d^{2}\beta }{\pi }\beta ^{n}\beta ^{\ast m}\exp \exp \left[
-|\beta |^{2}+\beta \alpha ^{\ast }+\beta ^{\ast }\alpha \right] =\left(
-i\right) ^{m+n}H_{m,n}\left( i\alpha ^{\ast },i\alpha \right) e^{|\alpha
|^{2}}  \label{48}
\end{equation}%
and the IWOP\ method that

\begin{eqnarray}
a^{n}\vdots e^{\lambda aa^{\dagger }}\vdots a^{\dagger m} &=&\int \frac{%
d^{2}z}{\pi }z^{n}e^{\lambda |z|^{2}}\left\vert z\right\rangle \left\langle
z\right\vert z^{\ast m}  \label{49} \\
&=&\int \frac{d^{2}z}{\pi }z^{n}z^{\ast m}:e^{-\left( 1-\lambda \right)
|z|^{2}+za^{\dagger }+z^{\ast }a-a^{\dagger }a}:  \notag \\
&=&\frac{1}{\left( 1-\lambda \right) ^{\left( n+m\right) /2+1}}\int \frac{%
d^{2}z}{\pi }z^{n}z^{\ast m}:e^{-|z|^{2}+\frac{1}{\sqrt{1-\lambda }}%
za^{\dagger }+\frac{1}{\sqrt{1-\lambda }}z^{\ast }a-a^{\dagger }a}:  \notag
\\
&=&\left( -i\right) ^{m+n}\left( 1-\lambda \right) ^{-\left( n+m\right)
/2-1}:e^{\lambda a^{\dagger }a/\left( 1-\lambda \right) }H_{m,n}(\frac{%
ia^{\dagger }}{\sqrt{1-\lambda }},\frac{ia}{\sqrt{1-\lambda }}):  \notag
\end{eqnarray}%
On the other hand (see the Appendix),
\begin{equation}
a^{n}\vdots e^{\lambda aa^{\dagger }}\vdots a^{\dagger m}=\sum_{l=0}^{\infty
}\frac{\lambda ^{l}}{l!}a^{l+n}a^{\dagger l+m}=\sum_{l=0}^{\infty }\frac{%
\lambda ^{l}}{l!}\left( -i\right) ^{m+n+2l}\colon H_{l+m,l+n}\left(
ia^{\dagger },ia\right) \colon  \label{50}
\end{equation}%
Comparing (\ref{49}) with (\ref{50}) we see

\begin{eqnarray}
&&\sum_{l=0}^{\infty }\frac{\lambda ^{l}}{l!}\left( -i\right)
^{m+n+2l}\colon H_{l+m,l+n}\left( ia^{\dagger },ia\right) \colon  \label{51}
\\
&=&\left( -i\right) ^{m+n}\left( 1-\lambda \right) ^{-\left( n+m\right)
/2-1}:e^{\lambda a^{\dagger }a}/^{\left( 1-\lambda \right) }H_{m,n}(\frac{%
ia^{\dagger }}{\sqrt{1-\lambda }},\frac{ia}{\sqrt{1-\lambda }}):  \notag
\end{eqnarray}%
noting $a^{\dagger }$ and $a$ are commutable within $:$ $:$, so replacing $%
ia^{\dagger }\rightarrow x,$ $ia\rightarrow y,$ $\lambda \rightarrow
-\lambda ,$ we obtain
\begin{equation}
\sum_{l=0}^{\infty }\frac{\lambda ^{l}}{l!}H_{l+m,l+n}\left( x,y\right)
=\left( 1+\lambda \right) ^{-\left( n+m\right) /2-1}e^{\lambda xy/\left(
1+\lambda \right) }H_{m,n}(\frac{x}{\sqrt{1+\lambda }},\frac{y}{\sqrt{%
1+\lambda }})  \label{52}
\end{equation}%
this is a new important generating function of $H_{l+m,l+n}\left( x,y\right)
.$ Especially, when $m=n$,

\begin{equation}
\sum_{l=0}^{\infty }\frac{\lambda ^{l}}{l!}H_{l+n,l+n}\left( x,y\right)
=\left( 1+\lambda \right) ^{-n-1}e^{\lambda xy/\left( 1+\lambda \right)
}H_{n,n}(\frac{x}{\sqrt{1+\lambda }},\frac{y}{\sqrt{1+\lambda }})  \label{53}
\end{equation}%
Using (\ref{44}) we can recast (\ref{53}) into%
\begin{equation}
\sum_{l=0}^{\infty }\frac{\left( n+l\right) !\left( -\lambda \right) ^{l}}{%
l!n!}L_{n+l}\left( z\right) =\left( 1+\lambda \right) ^{-n-1}e^{\lambda
z}/^{\left( 1+\lambda \right) }L_{n}\left( \frac{z}{1+\lambda }\right) .
\label{54}
\end{equation}%
This is the generalized negative binomial theorem involving Laguerre
polynomial. When $z=0$, it reduces to the negative-binomial formula%
\begin{equation}
\sum_{l=0}^{\infty }\frac{\left( n+l\right) !\left( -\lambda \right) ^{l}}{%
l!n!}=\left( 1+\lambda \right) ^{-n-1}  \label{55}
\end{equation}%
as expected.

\section{The normally ordered form of negative binomial state}

\bigskip when $n=m=s,$ Eq. (\ref{50}) reduces to

\begin{equation}
a^{s}\vdots e^{\lambda aa^{\dagger }}\vdots a^{\dagger s}=\left( -1\right)
^{s}\left( 1-\lambda \right) ^{-s-1}:e^{\lambda a^{\dagger }a/\left(
1-\lambda \right) }H_{s,s}(\frac{ia^{\dagger }}{\sqrt{1-\lambda }},\frac{ia}{%
\sqrt{1-\lambda }}):.  \label{56}
\end{equation}%
Taking $\lambda =\frac{\gamma }{\gamma -1},1-\lambda =\frac{1}{1-\gamma },%
\frac{1}{\gamma }-1=n_{c},$ then
\begin{eqnarray}
\rho _{s}\left( \gamma \right) &=&\frac{1}{s!\left( n_{c}\right) ^{s+1}}%
\vdots a^{s}e^{\frac{\gamma }{\gamma -1}aa^{\dagger }}a^{\dagger s}\vdots
\label{57} \\
&=&\frac{1}{s!\left( n_{c}\right) ^{s+1}}\left( -1\right) ^{s}\left(
1-\gamma \right) ^{s+1}:e^{\gamma a^{\dagger }a}H_{s,s}(i\sqrt{1-\gamma }%
a^{\dagger },i\sqrt{1-\gamma }a):  \notag \\
&=&\gamma ^{s+1}:e^{\gamma a^{\dagger }a}L_{s}(\left[ -\left( 1-\gamma
\right) a^{\dagger }a\right] :  \notag
\end{eqnarray}%
This is the normally ordered form of negative binomial state, which is
Laguerre polynomial weighted. We can re-derive Eq. (\ref{32}) by using Eq. (%
\ref{57}) and (\ref{9}), i.e.,

\begin{eqnarray}
1 &=&\frac{1-\gamma }{\gamma }\sum_{s=0}^{\infty }\rho _{s}\left( \gamma
\right) =\left( 1-\gamma \right) :e^{\gamma a^{\dagger }a}\sum_{s=0}^{\infty
}\gamma ^{s}L_{s}(\left[ -\left( 1-\gamma \right) a^{\dagger }a\right] :
\label{58} \\
&=&\left( 1-\gamma \right) :e^{\gamma a^{\dagger }a}\left( 1-\gamma \right)
^{-1}e^{\frac{-\gamma }{\gamma -1}\left( 1-\gamma \right) a^{\dagger }a}:=1.
\notag
\end{eqnarray}

In summary, we have found that the complete quantum state space can be
divided into mixed states in two ways, one is according to binomial states,
and another is according to negative binomial states. We now understand the
structure of Fock space more deeply. The new partition in negative binomial
states helps to build the generalized negative binomial theorem involving
Laguerre polynomial. The preparation possibility of these mixed states is
pointed out.

\section{Appendix}

We begin with presenting the operator identity

\begin{equation}
a^{n}a^{\dag m}=\left( -i\right) ^{m+n}:H_{m,n}\left( ia^{\dag },ia\right) :
\label{59}
\end{equation}%
where $:$ $:$ denoted normal ordering.

In fact, using the Baker-Hausdorff formula and Eq. (\ref{42}) we have

\begin{eqnarray}
e^{t^{\prime }a}e^{ta^{\dagger }} &=&e^{ta^{\dagger }}e^{t^{\prime
}a}e^{tt^{\prime }}=:e^{\left( -it^{\prime }\right) ia+\left( -it\right)
ia^{\dagger }-\left( -it\right) \left( -it^{\prime }\right) }:  \label{60} \\
&=&\sum_{m=0}\sum_{n=0}\frac{\left( -it\right) ^{m}\left( -it^{\prime
}\right) ^{n}}{n!m!}:H_{m,n}\left( ia^{\dag },ia\right) :.  \notag
\end{eqnarray}%
Comparing Eq. (\ref{60}) with

\begin{equation}
e^{t^{\prime }a}e^{ta^{\dagger }}=\sum_{m=0}\sum_{n=0}\frac{t^{\prime n}t^{m}%
}{n!m!}a^{n}a^{\dagger m},  \label{61}
\end{equation}%
we see%
\begin{equation}
a^{n}a^{\dag m}=\left( -i\right) ^{m+n}:H_{m,n}\left( ia^{\dag },ia\right) :.
\label{62}
\end{equation}%
Thus%
\begin{equation}
\sum_{l=0}\frac{\lambda ^{l}}{l!}\left( -i\right) ^{m+n+2l}\colon
H_{l+m,l+n}\left( ia^{\dagger },ia\right) \colon =\sum_{l=0}\frac{\lambda
^{l}}{l!}a^{l+n}a^{\dagger l+m}=a^{n}\vdots e^{\lambda aa^{\dagger }}\vdots
a^{\dagger m}.  \label{63}
\end{equation}

\end{document}